\documentclass[preprintnumbers,article,amsmath,amssymb,floatfix,10pt,prd,onecolumn,
superscriptaddress,nofootinbib]{revtex4-2}
\usepackage{bm}
\usepackage{amsfonts}
\usepackage{latexsym}
\usepackage[latin1]{inputenc}
\usepackage{graphicx}
\usepackage{amsmath}
\usepackage{palatino}
\usepackage{mathpazo}
\usepackage{textcomp}
\usepackage{float}
\usepackage{booktabs}
\usepackage{dcolumn}
\usepackage{ragged2e}
\usepackage{hyperref}
\hypersetup{colorlinks,citecolor=blue}
\hypersetup{colorlinks=true,linkcolor=red,filecolor=magenta, urlcolor=blue}
\usepackage{amsmath}
\usepackage{xcolor}
\usepackage{orcidlink}
\usepackage{epsfig}
\usepackage[caption=false]{subfig}
\usepackage{commath}
\usepackage{tabularx}
\usepackage{multirow}
\usepackage{mathtools}
\newcommand{\be}{\begin{equation}}
\newcommand{\ee}{\end{equation}}
\newcommand{\bea}{\begin{eqnarray}}
\newcommand{\eea}{\end{eqnarray}}
\newcommand{\eeas}{\end{eqnarray*}}
\newcommand{\beas}{\begin{eqnarray*}}

\def\jnl@style{\it}
\def\aaref@jnl#1{{\jnl@style#1}}

\def\aaref@jnl#1{{\jnl@style#1}}

\def\aj{\aaref@jnl{AJ}}                   
\def\apj{\aaref@jnl{ApJ}}                 
\def\apjl{\aaref@jnl{ApJ}}                
\def\apjs{\aaref@jnl{ApJS}}               
\def\apss{\aaref@jnl{Ap\&SS}}             
\def\aap{\aaref@jnl{A\&A}}                
\def\aapr{\aaref@jnl{A\&A~Rev.}}          
\def\aaps{\aaref@jnl{A\&AS}}              
\def\mnras{\aaref@jnl{Mon.~Not.~Roy.~Astron.~Soc.}}             
\def\prd{\aaref@jnl{Phys.~Rev.~D}}        
\def\prc{\aaref@jnl{Phys.~Rev.~C}}  
\def\prl{\aaref@jnl{Phys.~Rev.~Lett.}}    
\def\qjras{\aaref@jnl{QJRAS}}             
\def\skytel{\aaref@jnl{S\&T}}             
\def\ssr{\aaref@jnl{Space~Sci.~Rev.}}     
\def\zap{\aaref@jnl{ZAp}}                 
\def\nat{\aaref@jnl{Nature}}              
\def\aplett{\aaref@jnl{Astrophys.~Lett.}} 
\def\apspr{\aaref@jnl{Astrophys.~Space~Phys.~Res.}} 
\def\physrep{\aaref@jnl{Phys.~Rep.}}      
\def\physscr{\aaref@jnl{Phys.~Scr}}       
\def\commat{\aaref@jnl{Comm.~Math.~Phys.}}              
\def\science{\aaref@jnl{Science}}               
\def\cqg{\aaref@jnl{Classical Quant.~Grav.}}            
\def\jpcs{\aaref@jnl{JPCS}}                                     
\def\ijmpd{\aaref@jnl{Int.~J.~Mod.~Phys.~D}}                    
\def\grg{\aaref@jnl{Gen.~Relat.~Gravit.}}               
\def\rpp{\aaref@jnl{Rep.~Prog.~Phys.}}          
\def\npa{\aaref@jnl{Nucl.~Phys.~A}}        
\def\lrr{\aaref@jnl{Living Rev.~Rel.}}                   
\def\jcap{\aaref@jnl{J.~Cosmology Astropart.~Phys.}}    
\def\rmp{\aaref@jnl{Rev.~Mod.~Phys.}}   
\def\epjc{\aaref@jnl{Eur.~Phys.~J.~C}} 
\def\plb{\aaref@jnl{~Phy.~Lett.~B}} 
\def\mpla{\aaref@jnl{Mod.~Phy.~Lett.~A}} 
\def\arxiv{\aaref@jnl{arxiv.org}}

\allowdisplaybreaks[1]

\begin{document}
\color{black}       
\title{Cosmography in $f(Q,T)$ gravity with the specific $H(z)$ }
\author{S. H. Shekh\orcidlink{0000-0003-4545-1975}}
\email{da\_salim@rediff.com}
\affiliation{Department of Mathematics, S.P.M. Science and Gilani Arts, Commerce College, Ghatanji, Yavatmal, Maharashtra 445301, India}

\author{Anirudh Pradhan\orcidlink{0000-0002-1932-8431}}
\email{pradhan.anirudh@gmail.com}
\affiliation{Centre for Cosmology, Astrophysics and Space Science (CCASS), GLA University, Mathura-281 406, Uttar Pradesh, India}

\author{S. P. Gaikwad 
}
\email{gaikwad1.618@gmail.com}
\affiliation{Department of Mathematics, L. K. D. K. Banmeru Science College, Lonar, Maharashtra, India}

\author{K. R. Mule
}
\email{drkailasmule@gmail.com}
\affiliation{Department of Mathematics, S. D. M. Burungale Science and Arts College, Shegaon, Maharashtra, India.}
\begin{abstract}
\textbf{Abstract:} 

The dynamics of perfect fluid as a source in the context of modified gravity, specifically $f(Q,T)$ gravity, are examined within the Friedmann-Lemaître-Robertson-Walker (FLRW) cosmological model. This gravity is a generic function of the non-metricity scalar $Q$ and its trace $T$. We investigate the characteristics of the derived cosmological model using a parameterized form of Hubble's parameter, $H(z) = H_0\left[\Omega_{0m}(1+z)^3 + (1-\Omega_{0m})\right]^{\frac{1}{2}}$ (Mahmood et al. Int. J. Geom. Methods Mod. Phys., https://doi.org/10.1142/S0219887824502049). Our examination reveals how physical parameters such as energy density, pressure, and the equation of state parameter, among others, in our model accurately describe the physical behavior of the cosmos. Furthermore, we explore the kinematic parameters in our model, which provide valuable insights into the cosmos's expansion history, including its acceleration, deceleration, and the evolution of its large-scale structure. By exploring these aspects, we gain a deeper understanding of the cosmos's dynamics and evolution within the context of modified gravity.


\end{abstract}

\maketitle

\ 
\section{introduction}
Cosmological observations, such as Type Ia Supernovae \cite{1, 2}, cosmic microwave background radiation \cite{3, 4}, and large-scale structure \cite{5,6}, have consistently demonstrated that the universe's expansion is accelerating, with this phase having commenced in the relatively recent past. This phenomenon presents a significant challenge in modern cosmology, often attributed to the presence of dark energy, a mysterious component with negative pressure dominating the universe's evolution. \\
The equation of state parameter, $\omega = \frac{p}{\rho}$, is commonly employed to classify dark energy, with $\omega$ not necessarily being constant. Values of $\omega$ near -1 correspond to standard cosmology, while values slightly above -1 represent quintessence dark energy and those below -1 represent phantom dark energy. Constraints on $\omega$ have been derived using Type Ia supernovae (SNe-Ia) data, combined with other limits on the equation of state, yielding $-1.66 < \omega < -0.62$ and $-1.33 < \omega < -0.79$, respectively. The most recent constraint, obtained in 2009, limits $\omega$ to $-1.44 < \omega < -0.92$, based on a combination of cosmological data sets from CMB anisotropy, SNe-Ia luminosity distances, and galaxy clustering. Various studies have investigated the nature of dark energy, exploring alternative approaches, including modified matter sources and additional degrees of freedom in the action, such as quintessence energy, phantom energy, tachyon fields, k-essence, and generalized barotropic equations of state (e.g., Chaplygin gas and its modifications) \cite{7,8,9,10,11}. These proposals effectively reproduce the dynamical behavior of the cosmos.\\
Alternatively, modified gravity theories offer a theoretical framework that extends our understanding of cosmic evolution and structure by revising the fundamental laws of gravity as described by General Relativity. This introduction aims to provide an overview of modified gravity's concept and significance in cosmology. While General Relativity, formulated by Albert Einstein in the early 20th century, has been the cornerstone of modern cosmology, unresolved mysteries and observational anomalies, such as dark matter and dark energy, cosmic microwave background radiation, and large-scale structure, have prompted cosmologists to explore modified gravity theories. These theories propose modifications to General Relativity's equations to address the aforementioned cosmological challenges, offering a potential solution to the observed phenomena. Recent decades have witnessed the emergence of various concepts in the scientific literature aimed at addressing the persisting cosmological challenges. Among these, modified theories of gravity have garnered significant attention as a promising approach to resolving the outstanding issues in cosmology, particularly the enigmatic nature of dark energy (DE) and dark matter (DM). The modified gravity framework offers a viable alternative to traditional cosmological models, providing a new paradigm for understanding the universe's evolution and structure, and potentially alleviating the shortcomings associated with DE and DM. Motivated by the limitations of traditional General Relativity (GR), researchers have explored alternative theories to address the existing cosmological challenges. Various modified gravity theories have been proposed in the literature, including: $f(R)$ gravity \cite{12}, $f(T)$ gravity \cite{13,14,15}, $f(T, B)$ gravity \cite{16}, $f(R, T)$ gravity \cite{17,18}, $f(G)$ gravity \cite{19}, $f(R, G)$ gravity \cite{20} etc. are the theories aim to extend or modify the traditional GR framework, providing new insights into the universe's evolution and structure, and potentially resolving the longstanding issues associated with dark energy and dark matter.\\
Currently, a plethora of modified gravity models have been proposed and are emerging in the literature, aiming to address the cosmological challenges associated with dark energy and dark matter. To ascertain their viability in describing the universe's dark sector, rigorous testing and evaluation are essential. Notably, a novel alternative, $f(Q)$ gravity, has recently been introduced, offering a geometric interpretation incorporating a non-metricity term $Q$ \cite{21,22}. This term, defined as $Q_{\gamma \mu \nu } = \nabla {\gamma }g{\mu \nu }$, describes the variation in the measurement of a vector during parallel transport, providing a new framework for understanding the universe's evolution and structure. Recent studies on $f(Q)$ gravity have been reported in Ref. \cite{23,24}. A particularly intriguing extension of this framework involves the coupling of non-metricity $Q$ with the matter sector, specifically through the trace of the energy-momentum tensor, $T$. This modified theory, denoted as $f(Q, T)$ gravity, can be formulated by introducing an arbitrary function of both $Q$ and $T$ in the action, as demonstrated in Ref. \cite{25}. Specifically, Ref. \cite{25} presents the modified Einstein-Hilbert action for extended symmetric teleparallel gravity, which incorporates this $f(Q, T)$ coupling as,
\begin{equation}\label{e1}
	S=\int \left[ \frac{1}{16\pi }f(Q,T)+L_{m}\right] \sqrt{-g}d^{4}x,
\end{equation}%
where $g$ is the metric tensor's determinant $g_{\mu \nu }$ i.e. $g=\det \left( g_{\mu \nu }\right) $, $f(Q,T)$ is the generic function of the non-metricity scalar $Q$ and its trace, $T$ and $L_{m}$ is the usual matter Lagrangian.\\ 
The non-metricity scalar $Q$ is
defined as
\begin{equation}\label{e2}
	Q\equiv -g^{\mu \nu }\left( L^{\delta }{}_{\alpha \mu }L^{\alpha }{}_{\nu
		\delta }-L^{\delta }{}_{\alpha \delta }L^{\alpha }{}_{\mu \nu }\right) ,
\end{equation}%
where $L^{\delta }{}_{\alpha \gamma }=-\frac{1}{2}g^{\delta \eta }\left( \nabla
_{\gamma }g_{\alpha \eta }+\nabla _{\alpha }g_{\eta \gamma }-\nabla _{\eta
}g_{\alpha \gamma }\right)$, $Q_{\gamma \mu \nu }=\nabla _{\gamma }g_{\mu \nu }$ and 
$Q_{\delta }=g^{\mu \nu }Q_{\delta \mu \nu },\text{ \ \ \ \ }\widetilde{Q}%
_{\delta }=g^{\mu \nu }Q_{\mu \delta \nu }$ are respective the dis-formation tensor, non-metricity tensor, and the trace of the non-metricity tensor.\\
In addition, the field equations of $f\left( Q,T\right) $ gravity are obtained by modifying the action, Eq. (\ref{e1}), in relation to the metric tensor $g_{\mu \nu }$
\begin{widetext}
	\begin{equation}\label{e6}
		-\frac{2}{\sqrt{-g}}\nabla _{\delta }\left( f_{Q}\sqrt{-g}P^{\delta }{}_{\mu
			\nu }\right) -\frac{1}{2}fg_{\mu \nu }+f_{T}\left( T_{\mu \nu }+\theta _{\mu
			\nu }\right) -f_{Q}\left( P_{\mu \delta \alpha }Q_{\nu }{}^{\delta \alpha
		}-2Q^{\delta \alpha }{}_{\mu }P_{\delta \alpha \nu }\right) =8\pi T_{\mu \nu
		},
	\end{equation}
\end{widetext}
where $f_{Q}=\frac{df\left( Q,T\right) }{dQ}$, $f_{T}=\frac{df\left(
	Q,T\right) }{dT}$, and the covariant derivative is denoted by the symbol $\nabla _{\delta }$. According to Eq. (\ref{e6}), the tensor $\theta _{\mu \nu }$ affects how the field equations of $f(Q,T)$ extended symmetric teleparallel gravity function. \\
Subsequently, Ref. \cite{25} employed the cosmological evolution equations for a flat, homogeneous, and isotropic universe, which generalize the Friedmann equations of standard General Relativity, to investigate the cosmological implications of this modified theory for specific functional forms of $f(Q, T)$. Notably, these functional forms feature an additive combination of $Q$ and $T$, enabling an examination of the theory's cosmological consequences in a straightforward and analytically tractable manner. Recent studies have revealed that this modified gravity theory significantly impacts the nature of gravitational influence and the equation of motion in the Newtonian limit \cite{26}. Furthermore, the cosmological linear perturbation theory of $f(Q, T)$ gravity, developed by Najera and Fajardo \cite{27}, suggests that the theory's predictions may be testable using Cosmic Microwave Background (CMB) and standard siren data. Additionally, the energy conditions and limitations on various forms of $f(Q, T)$ gravity have been investigated in \cite{28}, utilizing the deceleration parameter, Hubble parameter, and consistency with the  $f(Q, T)$ gravity model has been proposed in \cite{29}, wherein the vector field $W_{\mu}$ entirely characterizes the scalar non-metricity, offering a new perspective on this modified gravity framework.\\
Motivated by the studies discussed in the aforementioned references, this work presents a cosmological model that describes the large-scale evolution of the Universe. Within the framework of the isotropic and spatially homogeneous Friedmann-Lemaître-Robertson-Walker (FLRW) Universe, the author considers a perfect fluid matter source and explores three modified gravity models: $f(Q,T)=\alpha Q+ \beta$, $f(Q,T)=\alpha Q^{(n+1)}+ \beta T$, and $f(Q,T)=\alpha Q+ \beta Q^2+\gamma T$. The validity of these models will be examined through the energy conditions, providing insight into the viability of the proposed cosmological scenario. This paper is structured as follows: Section 2 derives the field equations within the isotropic framework. Section 5 examines the viability of a specific $f(T)$ gravity model. Finally, Section 6 presents the concluding remarks, summarizing the key findings and implications of this study.

\section{Field equations in isotropic framework}

To facilitate the solution of field equations in $f(Q,T)$ extended symmetric teleparallel gravity, simplifying assumptions are often necessary. In this work, we adopt the homogeneous, isotropic, and spatially flat Friedmann-Robertson-Walker (FRW) metric, given by:

\begin{equation}\label{e8}
	ds^{2}=-dt^{2}+\delta_{ij} g_{ij} dx^{i} dx^{j},{\;\;\;\;} i,j=1,2,3,.....N,
\end{equation}
This choice of metric allows us to explore the cosmological implications of $f(Q,T)$ gravity in a straightforward and analytically tractable manner.
where $g_{ij}$ are the function of $(-t, x^{1}, x^{2}, x^{3})$ and $t$ refers to the cosmological/cosmic time measure in Gyr. In the four-dimensional FRW space-time, the equation above yields the following:
\begin{equation}\label{e9}
	\delta_{ij} g_{ij}=a^{2}(t,x)
\end{equation} 
where $a$ be the average scale factor of the Universe and $t$ is the cosmic time in Gyr. The aforementioned relationships demonstrate that all three metrics are equivalent in the FRW universe (i.e $g_{11} = g_{22} = g_{33} =a^{2}(t, x)$). 
Consequently, the non-metricity scalar for the line element in  Eq. (\ref{e8}) is denoted by $Q=6H^{2}$, where $H$ be the average Hubble's parameter of the form $H=\frac{\dot{a}}{a}$.\\
We consider a perfect fluid for which: 
\begin{equation}\label{e13}
T_{\nu }^{\mu }=diag\left( -\rho ,p,p,p\right) , 
\end{equation}%
where $\rho $ is the energy density and the isotropic pressure is $p$. The field equations (\ref{e6}) for the metric
(\ref{e8}) yield
\begin{equation}\label{e14}
\kappa^{2} \rho =\frac{f}{2}-6FH^{2}-\frac{2\widetilde{G}}{1+\widetilde{G}}\left( 
\overset{.}{F}H+F\overset{.}{H}\right) ,  
\end{equation}
\begin{equation}\label{e15}
\kappa^{2} p=-\frac{f}{2}+6FH^{2}+2\left( \overset{.}{F}H+F\overset{.}{H}\right) .
\end{equation}
where, where $\kappa^{2} \widetilde{G}\equiv f_{T}$ (here $\kappa^{2}=1$) and $F\equiv f_{Q}$  are derivatives with respect to  $T$ and $Q$,  respectively. The Hubble parameter $H$ is given by $H\equiv \dot{a}/{a}$ and $\left( \text{\textperiodcentered }\right) $ is $d/dt$. \\
Numerous physical parameters or attributes within a cosmology are closely associated with the above energy density and isotropic pressure, and their behavior may frequently be studied by analyzing their expressions or interpreting their graphical representations. The following will be the examination of the expressions of certain important components, such as the equation of state parameter, and the energy conditions.\\
\textit{The Equation of State parameter is}
\begin{equation}\label{e16}
\omega=\frac{p}{\rho}=-1+\left(\frac{1}{\kappa^{2} \rho}\right) \left(\frac{2 \kappa^{2}+f_{T}}{\kappa^{2} +f_{T}}\right)\left( \overset{.}{F}H+F\overset{.}{H}\right).   
\end{equation}

Interestingly, the final nature of DE is often classified using the so-called Equation of State (EoS) parameter, which quantifies the correlation between spatially homogeneous pressure and energy density. We may now comprehend the importance of the EoS parameter, $\omega < -1/3$, which is necessary for rapid cosmic expansion according to recent cosmological investigations. The most significant options in this classification are scalar field models with an EoS value of $-1<\omega <-1/3$, also referred to as a Quintessence field DE \cite{30,31,32}, as opposed to $\omega <-1$, which is called a phantom field DE \cite{33}. Additionally, the EoS parameter for DE is now $\omega_0 = -1.084 \pm 0.063$ according to the combined observations of WMAP9 and the $H_0$ measurements, Supernovae of type Ia (SNe Ia), CMB, and BAO (Baryon Acoustic Oscillations). Also, we improved $\omega_0 = -1.028^{+0.032}_{-0.032}$ in 2018 and the Planck collaboration found in 2015 that $\omega_0 = -1.006^{+0.0451}_{-0.0451}$ \cite{34,35}. \\

\section{ Specific $H(z)$ and some Cosmographic  parameters}

Within the framework of symmetric teleparallel gravity, we have assumed a perfect fluid as the matter content of the Universe. To determine the expansion rate, we define the dimensionless function $E(z)$ as:
$$E(z) = \frac{H^2(z)}{H_0^2} = \Omega_{0m}(1+z)^3 + \Omega_{0_{Q,T}}$$
where $H(z)$ is the Hubble parameter at redshift $z$, $H_0$ is the present-day value of the Hubble constant, and $\Omega_{0m}$ is the present-day value of the matter energy density parameter. Note that $\Omega_{0_{Q,T}}$ represents the energy density parameter arising from the geometry of $f(Q,T)$ gravity. An alternative functional form of $E(z)$ can be expressed as   \cite{36} $$	E(z) = \Omega_{0m}(1+z)^{3} + \alpha(1+z)^{2} + \beta(1+z) + \mu   $$
It is important to note that at $z = 0$, the Hubble parameter $H(z)$ equals the present-day value $H_0$, which implies that $E(z) = 1$ at $z = 0$. This constraint validates above Eq. $\alpha$, $\beta$, and $\mu$ to $\alpha + \beta + \mu = 1 - \Omega_{0m}$. To satisfy this condition, we consider the simplest functional form as $\Omega_{0_{Q,T}} = 1 - \Omega_{0m}$. Consequently, we can rewrite the above equation as:
\begin{equation}
H(z) = H_0\left[\Omega_{0m}(1+z)^3 + (1-\Omega_{0m})\right]^{\frac{1}{2}}
\end{equation}
This expression represents the Hubble parameter $H(z)$ in terms of the redshift $z$, the present-day matter energy density parameter $\Omega_{0m}$, and the present-day Hubble constant $H_0$.

\begin{figure}[H]
		\centering
		\includegraphics[scale=0.6]{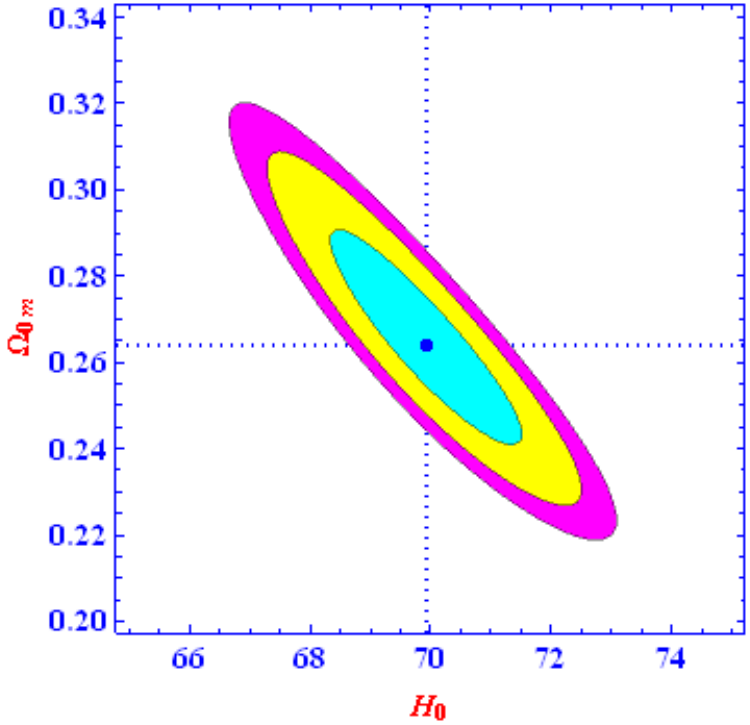}
		\caption{Above figure shows the combined visualization of two dimensional contours at 1$\sigma$ \& 2$\sigma$ confidence regions by bounding ourmodel with OHD data sets and OHD + Pantheon compilation of SN Ia data}\label{HP}
\end{figure}

In Lohakare et al. \cite{36}, the authors have analyzed the geometrical and dynamical parameters of modified teleparallel - Gauss - Bonnet model and constrained $\Omega_{0m}$ and other constants by fitting the model with observation Hubble data and Pantheon compilation of SN Ia data.Also, the same author have considered $H _{0} = 70.7$ km/s/Mpc. In this research, we have constrained both $H_{0}$ and $\Omega_{0m}$ by fitting the experimental data and their measurements. Furthermore, we also regenerate a symmetric teleparallel gravity theory by employing energy conditions. Very recently, Mahmood et al. \cite{36a} studied how modified gravity affects the universe on a large scale, focusing on a specific type of modified gravity called $f(Q)$ cosmology. They used energy conditions to create different models, considering a universe that's speeding up, a mysterious substance called quintessence, and a constant called Lambda. Using different sets of observational data, including supernovae and cosmic chronometer data, they estimated the Hubble constant ($H_0$) to be around $70.37^{+0.84}_{?0.92}$ from $H(z)$ data and $70.02^{+0.44}_{?0.25}$ from pantheon compilation of SN Ia data whereas respectively the value of matter energy density parameter ($\Omega_{0m}$) to be around $0.260^{+015}_{?0.010}$ (OHD) and $0.270^{+025}_{?0.014}$ (SN Ia). In this study, we combine the data sets used by Irfan et al. to constrain the values of the parameters. By combining the Supernova Pantheon sample, cosmic chronometer data, and other observational data, we aim to obtain more precise estimates of the parameters. This combined analysis allows us to better understand the properties of the universe and the modified gravity model, $f(Q,T)$ cosmology. By constraining the parameter values, we can gain insights into the universe's evolution, expansion, and the nature of dark energy. The below Fig. \ref{HP}, shows the combined visualization of two-dimensional contours, which represent the confidence regions of our model's parameters. The contours are based on two different data sets: OHD (cosmic chronometer data) and OHD + Pantheon (a combination of cosmic chronometer data and supernovae data). The contours are shown at two confidence levels: 1$\sigma$ (68\% confidence) and 2$\sigma$ (95\% confidence). This visualization helps us understand the constraints on our model's parameters and how well they fit the data. The contours provide a visual representation of the uncertainty and accuracy of our model's predictions.

Our analysis combines two data sets: cosmic chronometer data (OHD) and a combination of cosmic chronometer data and supernovae data (OHD + Pantheon). From this combined analysis, we estimate the values of two important cosmic parameters:
\begin{itemize}
	\item - Hubble constant ($H_0$): 69.94
	\item- Matter energy density parameter ($\Omega_{0m}$): 0.264
\end{itemize}
These values represent the best fit to the data and provide insights into the expansion history and composition of the universe.
The dynamics of a cosmological model offer valuable insights into the behavior and evolution of the Universe. In this study, we utilize some complementary approaches to investigate the dynamics of our model: (i) the deceleration parameter (ii) the state-finder parameters, (ii) the $O_m(z)$-diagnostics. These approaches enable us to examine the comprehensive understanding of the model's cosmological implications.

\subsubsection{Deceleration parameter}
The transition from a decelerating to an accelerating phase is a pivotal aspect of understanding the universe's dynamics. Initially, the universe's expansion was slowing down due to the strong gravitational attraction between matter and radiation. However, as the universe expanded and matter became more dispersed, the gravitational force weakened, altering the cosmic dynamics. The universe has now entered a phase of cosmic acceleration, characterized by a negative deceleration parameter ($q$). Analyzing this transition is crucial for understanding the underlying mechanisms driving the universe's expansion. The deceleration parameter provides a quantitative measure of this acceleration, allowing us to assess the current state of cosmic expansion and gain insights into the universe's evolution. Hence, The deceleration parameter ($q$), defined in terms of the Hubble parameter ($H$) as,
\begin{equation}
	q(z) = -1+\frac{(1+z)}{H(z)}\frac{dH(z)}{dz}.
\end{equation}

The deceleration parameter ($q$) serving as a diagnostic tool. A positive value of $q$ indicates a decelerating phase, while a negative value signifies an accelerating phase. Recent cosmological observations have constrained the present-day value of $q$ to $q_0 = -0.51^{+0.09}_{-0.01}$ and $q_0 = -0.5422_{-0.0826}^{+0.0718}$. Furthermore, the transition redshift from deceleration to acceleration has been measured to be $z_t = 0.65^{+0.19}_{-0.17}$ and $z_t = 0.8596_{-0.2722}^{+2886}$. These estimates provide valuable insights into the evolution of the Universe's expansion \cite{37,38,39}.

\subsubsection{State-finder parameters}
The significance of dark energy in driving the accelerating cosmic expansion is well-established. In recent decades, there has been a surge of interest in elucidating the origin and fundamental properties of dark energy, leading to the development of numerous dark energy models. To discern between these models, both quantitatively and qualitatively, Sahni et al. \cite{40} introduced a statefinder diagnostic tool. This approach employs a pair of geometrical parameters, known as statefinder parameters ($r$, $s$), which are defined as:
 \begin{eqnarray}
	j(z) &=& q(z) + 2q^{2}(z) + (1+z)\frac{dq(z)}{dz},\label{eq:27} \\
	s(z) &=& \frac{j(z)-1}{3\left(q(z)-\frac{1}{2}\right)}, ~~~~~~~~~~~~~~\left( q\neq \frac{1}{2}\right).
	\label{eq:28}
\end{eqnarray}

These parameters provide a robust means of distinguishing between various dark energy models, enabling a deeper understanding of the underlying mechanisms driving the cosmic acceleration. The authors present in Ref. \cite{40,41,42} have categorized various dark energy models based on the state-finder parameters such as
\begin{itemize}
	\item \( (j=1,~s=0) \rightarrow \Lambda \)CDM;
	\item \( (j<1,~s>0) \rightarrow \) Quintessence;
	\item \( (j>1,~s<0) \rightarrow \) Chaplygin Gas;
	\item \( (j=1,~s=1) \rightarrow \) SCDM.
\end{itemize}
Notably, the point $\{j,~s\} = \{1, 0\}$ serves as a reference point, representing the flat $\Lambda$CDM model. By using this as a benchmark, one can evaluate the deviation of other models from the flat $\Lambda$CDM model, providing a quantitative measure of their distinction. In the statefinder plane, the sign of the parameter $s$ serves as a discriminator between quintessence-like and phantom-like dark energy (DE) models. Specifically, positive values of $s$ indicate quintessence-like behavior, whereas negative values of $s$ signify phantom-like behavior. Furthermore, the trajectory of the statefinder pair $\{j, s\}$ passing through the point $\{1, 0\}$ in the $j-s$ plane marks a transition from phantom to quintessence behavior, highlighting a shift in the DE dynamics. This underscores the utility of the statefinder parameters as a robust tool for categorizing and distinguishing between various DE models, as demonstrated in \cite{43}.

\subsubsection{$O_m(z)$ Diagnostic:}

The $O_m(z)$ diagnostic has been established as a versatile tool for probing the accelerated expansion of the Universe, offering an alternative approach to traditional methods. This diagnostic is capable of distinguishing a broad range of dark energy models, including quintessence, phantom, and  $O_m(z)$ diagnostic exhibits high sensitivity to the equation of state (EoS) parameter, as demonstrated by numerous studies in the literature \cite{44,45,46}. The $O_m(z)$ diagnostic is defined as,

\begin{equation}
	O_m(z) = \frac{\left[\frac{H(z)}{H_{0}}\right]^2 - 1}{(1+z)^{3}-1}, \hspace{0.5cm}
\end{equation}

Notably, the slope of the $O_m(z)$ function serves as a discriminator between different dark energy models: a positive slope is indicative of a phantom phase with $\omega < -1$, whereas a negative slope characterizes a quintessence region with $\omega > -1$. This property enables the $O_m(z)$ diagnostic to provide valuable insights into the nature of dark energy.

\section{Physical aspects and analysis of the models}
In this section, we delve into the physical parameters that underpin the model, exploring their significance and impact on the cosmological evolution. By examining these parameters, we can uncover the fundamental characteristics of the model and elucidate its physical implications.
\subsubsection{Energy conditions}
The evolution of the universe is intricately linked to the cosmological parameters, particularly the equation of state (EoS), as discussed earlier. However, the Raychaudhuri equation provides a set of energy conditions that play a crucial role in predicting cosmic acceleration in modern cosmology. These energy conditions are fundamental to General Relativity (GR), as they establish the existence theorems for black holes and space-time singularities \cite{48}. Various authors have contributed to the study of energy conditions, which can take several forms, including the null energy condition (NEC), dominant energy condition (DEC), and strong energy condition (SEC). In this work, we will examine these well-known energy conditions to verify the viability of the model in explaining cosmic acceleration.\cite{49}

\noindent
NEC: \begin{itemize}
	\item For each null vector, $T_{ij}u^{i}u^{j} \geq 0 \Rightarrow \rho+ p \geq 0.$
\end{itemize}
DEC: \begin{itemize}
	\item for any time like vector: $
	T_{ij}u^{i}u^{j} \geq 0 \Rightarrow \rho- p\geq 0~~\text{and}~~ T_{ij}u^{j}$~\text{not~spacelike}.
\end{itemize}
SEC: 
\begin{itemize}
	\item for any time-like vector: $\left(T_{ij}-\frac{1}{2}Tg_{ij}\right)u^{i}u^{j} \geq 0 \Rightarrow  \rho + 3p \geq 0$.
\end{itemize}
Next, we explore various functional forms of $f(Q,T)$ gravity models, as presented in the literature \cite{50}. These models represent extensions of the original $f(Q)$ gravity theory, incorporating the trace of the energy-momentum tensor $T$ to potentially capture additional aspects of gravitational physics.

\begin{itemize}
    \item  (a) $f(Q,T)=\alpha Q+ \beta T$
    \item (b)  $f(Q,T)=\alpha Q^{(n+1)}+ \beta T$
    \item (c) $f(Q,T)=\alpha Q+ \beta Q^2+\gamma T$
\end{itemize}

 
 
 The modified field equations, denoted by equations (\ref{e14}) and (\ref{e15}), are the two independent equations that characterize the dynamics of the model. Three unknowns were employed in these underlying formulations. This implies that in order to completely explain and solve this system, we need one more equation. Experts in the subject have responded to this by using a widely accepted approach called the model independent way study of model, which often considers a scheme of parameterization of a cosmological parameter. Nonetheless, based on how certain geometrical parameters are parametrized, several kinds of intriguing dark energy and modified gravity models are accessible \cite{51,52,53,54,55}.


\subsection{model-I}
We initiate our investigation by exploring the inaugural model of $f(Q,T)$ gravity, a pioneering framework that extends the boundaries of general relativity by incorporating the non-metricity scalar $Q$ and the trace of the energy-momentum tensor $T$. Specifically, this foundational model posits a linear functional form of $f(Q,T)=\alpha Q+ \beta T$, where $\alpha$ and $\beta$ are dimensionless constants that regulate the strength of the non-metricity and matter-energy interactions, respectively. The term $\alpha Q$ encapsulates the effects of non-metricity, which quantifies the deviation from the Riemannian geometry, while the term $\beta T$ represents the contribution from the matter-energy sector, encompassing the dynamics of cosmic fluids. By scrutinizing this elementary model, we can uncover the fundamental implications of non-metricity and its interplay with matter-energy density, providing a springboard for understanding the more complex dynamics of the universe.\\
For the aforementioned $f(Q,T)$ model, we derive the expressions for the cosmographic parameters and discuss their implications, based on the components of the field equations given in equations (\ref{e14}) - (\ref{e15}). We then proceed to discuss the cosmological implications of these parameters, examining the dynamics of the universe 
\\
\textit{\textbf{The energy density  is}}

\begin{equation}
\rho =\frac{\alpha  \left(\beta  \dot{H}-3 (\beta +1) H^2\right)}{2 \beta ^2+3 \beta +1}
\end{equation}

\begin{figure}[H]
	\begin{minipage}{0.49\textwidth}
		\centering
		\includegraphics[scale=0.7]{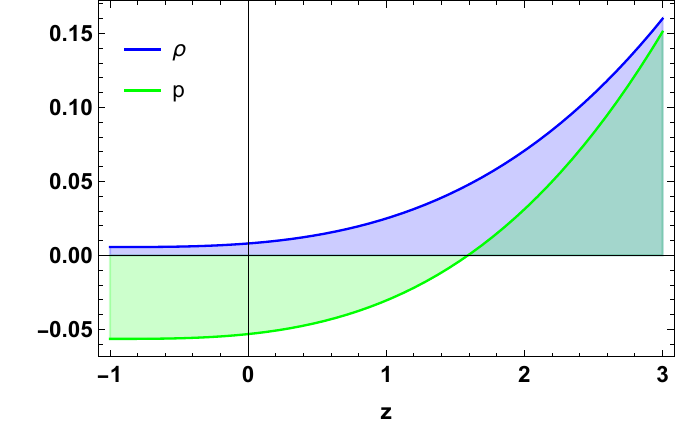}
		\caption{Above figure shows the behavior of $\rho$ and $p$ versus $z$ with the  constraint values of the cosmological free parameters limited by $H(z)$ data set.}\label{M1denp}
	\end{minipage}\hfill
	\begin{minipage}{0.49\textwidth}
		\centering
		\includegraphics[scale=0.6]{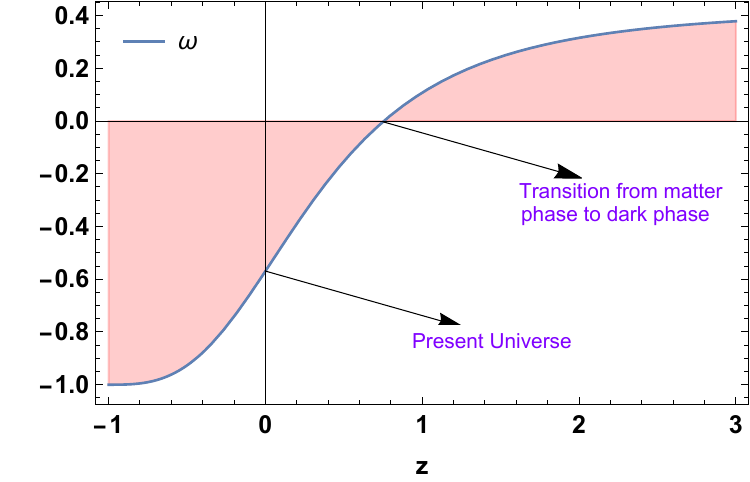}
		\caption{Above figure shows the behavior of $\omega$ versus $z$ with the  constraint values of the cosmological free parameters limited by $H(z)$ data set.}\label{M1EOS}
	\end{minipage}
\end{figure}

The energy density of the universe in the current $f(Q,T)$ gravity model is given by equation (36), and its geometrical behavior with respect to redshift $z$ is illustrated in Fig. \ref{M1denp} (blue curve). $\rho$ is a monotonically increasing function of redshift, diverging to infinity as $z$ approaches infinity ($\rho \rightarrow \infty$ as $z \rightarrow \infty$). This indicates that the universe was extremely dense in its early stages. Furthermore, as $z$ approaches -1, the energy density tends towards a small positive value, suggesting that the universe is undergoing expansion. These results provide valuable insights into the cosmological evolution of the universe within the framework of this $f(Q,T)$ gravity model.\\

\textit{\textbf{The isotropic pressure is}}
\begin{equation}
p = \frac{\alpha  \dot{H}}{\beta +1}+\frac{\alpha  \left(\dot{H}+3 H^2\right)}{2 \beta +1}
\end{equation}
Astronomical observations suggest that dark energy is responsible for the accelerating expansion of the universe, with negative pressure being a key indicator of its presence. Our model's universe pressure, $p$, as a function of redshift $z$, is depicted in Fig. \ref{M1denp} (green curve). This plot reveals that pressure increases monotonically with redshift, diverging to infinity as $z$ approaches infinity ($p \rightarrow \infty$ as $z \rightarrow \infty$). Notably, pressure becomes negative in the present era ($z=0$) and remains so in the future ($z<0$), consistent with the expected behavior of dark energy. These findings provide further evidence for the validity of our $f(Q,T)$ gravity model in describing the universe's accelerating expansion.

\textit{\textbf{The equation of state parameter is}}

\begin{equation}
\omega = -1+\frac{(\beta +2) (2 \beta +1) \dot{H}}{\beta  \dot{H}-3 (\beta +1) H^2}
\end{equation}
The evolution of the equation of state (EoS) parameter $\omega$ with redshift $z$ is illustrated in Fig. \ref{M1EOS} ($\omega>0$), transitions to a quintessence phase ($\omega>-1$) at present, and asymptotically approaches the ($\omega=-1$) in the future. This evolutionary trajectory implies a transition from a decelerating to an accelerating phase of cosmic expansion. Notably, the present-day value of the EoS parameter, $\omega_0 = -0.58318$, confirms that the universe is currently undergoing acceleration. These findings provide valuable insights into the dynamical evolution of the universe within the framework of our $f(Q,T)$ gravity model.\\

\begin{figure}[H]
		\centering
		\includegraphics[scale=0.7]{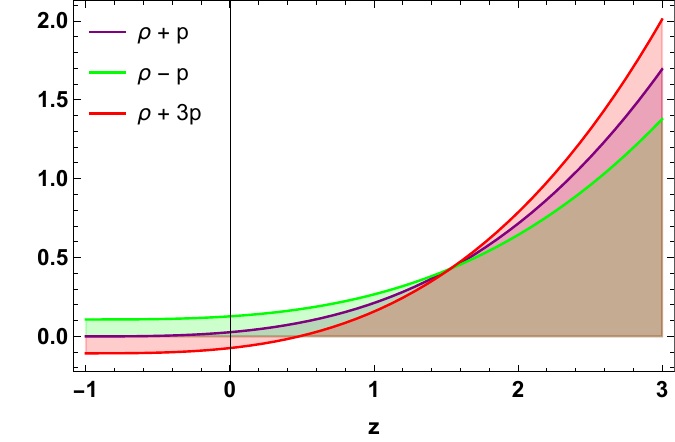}
		\caption{Above figure shows the behavior of $\rho + p, \rho - p$ and $\rho + 3p$ versus $z$ with the  constraint values of the cosmological free parameters limited by $H(z)$ data set.}\label{M1EC}
\end{figure}

\textit{\textbf{The energy conditions are}}\\
The expressions of Null, Dominant and strong energy conditions are obtained as
\begin{small}
	\begin{equation}\label{eq:27}
	\rho + p = -\frac{3 \alpha  H_0^2 \Omega_{0m} (z+1)^3}{\beta +1},
	\end{equation}
		\begin{equation}\label{eq:27}
		\rho - p = -\frac{3 \alpha  H_0^2 (\Omega_{0m} (z (z (z+3)+3)-1)+2)}{2 \beta +1},
	\end{equation}
	
		\begin{equation}\label{eq:27}
		\rho +3 p = \frac{3 \alpha  H_0^2 (2 (\beta +1)-\Omega_{0m} (5 \beta +(3 \beta +1) z (z (z+3)+3)+3))}{(\beta +1) (2 \beta +1)}.,
	\end{equation}

\end{small}
The dynamical evolution of the energy conditions is illustrated in Fig. \ref{M1EC}, utilizing the constrained values of the free parameters. Our analysis reveals that the Null Energy Condition (NEC) exhibits a decreasing trend during the early universe, remaining positive throughout but ultimately vanishing at late times. In contrast, the Dominant Energy Condition (DEC) remains consistently positive, without violation, throughout the cosmological evolution. Meanwhile, the Strong Energy Condition (SEC) exhibits no violation during the early universe, but does violate at late times. These findings underscore the dynamic and epoch-dependent nature of the energy conditions, highlighting their evolving role in shaping the cosmological landscape.

\subsection{model-II}
\begin{figure}[H]
		\centering
		\includegraphics[scale=0.6]{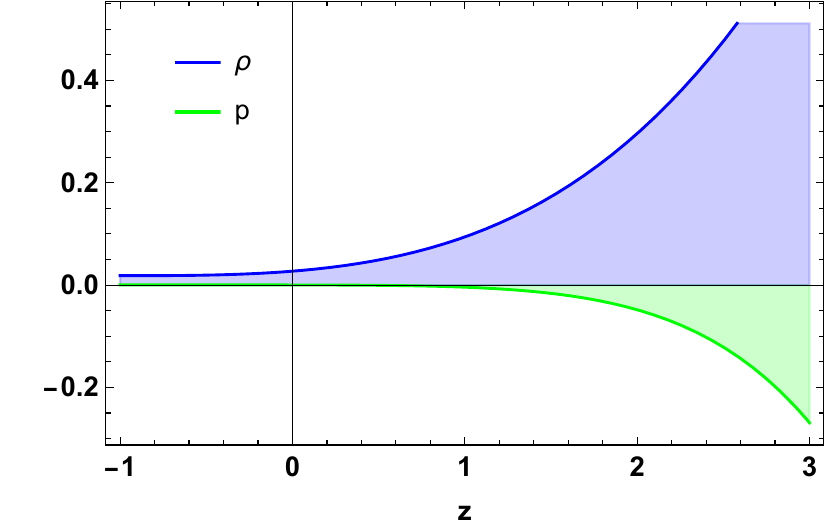}
		\caption{Above figure shows the behavior of $\rho$ and $p$ versus $z$ with the  constraint values of the cosmological free parameters limited by $H(z)$ data set.}\label{M2denp}
\end{figure}

Next, we delve into the $f(Q,T)$ gravity model, a modified theoretical framework that extends the realm of general relativity by incorporating the non-metricity scalar $Q$ and the trace of the energy-momentum tensor $T$. Specifically, this model postulates a functional form of $f(Q,T)=\alpha Q^{(n+1)}+ \beta T$, where $\alpha$ and $\beta$ are dimensionless constants, and $n$ is a real exponent that characterizes the power-law dependence on $Q$. This expression encapsulates a rich phenomenology, enabling us to explore the interplay between non-metricity and matter-energy density. The first term, $\alpha Q^{(n+1)}$, captures the effects of non-metricity, while the second term, $\beta T$, represents the contribution from the matter-energy sector. By investigating this model, we can gain insights into the potential deviations from standard general relativity and the implications for cosmological dynamics.\\
Here's a polished version of the paragraph with more detailed descriptions:

For the aforementioned $f(Q,T)$ model, we embark on a comprehensive analysis by deriving the explicit expressions for the cosmographic parameters which are intricately linked to the components of the field equations, as presented in equations (\ref{e14}) and (\ref{e15}), which form the foundation of our investigation. \\
\textit{\textbf{The energy density  is}}
\begin{equation}
\rho =-\frac{\alpha  6^n H^{2n} \left(3 (\beta +1) H^2 (2 n+1)-\beta  (n+1) (\dot{H}+2 \dot{H} n)\right)}{(\beta +1) (2 \beta +1)}
\end{equation}
The energy density, $\rho$, in the current model exhibits a identical behavior to that of model-I (see fig. \ref{M2denp} (blue curve)), characterized by a monotonically increasing function of redshift, which diverges to infinity as $z$ approaches infinity ($\rho \rightarrow \infty$ as $z \rightarrow \infty$). This implies a extremely high density universe in its early stages. Additionally, as $z$ approaches -1, the energy density approaches a small positive value, indicating an expanding universe. These findings offer significant insights into the cosmological evolution of the universe within the context of this $f(Q,T)$ gravity model, providing a consistent picture of the universe's evolution across different stages.\\
\textit{\textbf{The isotropic pressure is}}
\begin{widetext}
\begin{equation}
	p = \frac{\alpha  6^n H^{2n} \left((3 \beta +2) \dot{H} (n+1)+3 (\beta +1) H^2 (2 n+1)+2 (3 \beta +2) \dot{H} n (n+1)\right)}{(\beta +1) (2 \beta +1)}
\end{equation}
\end{widetext}

\begin{figure}[H]
	\centering
	\includegraphics[scale=0.6]{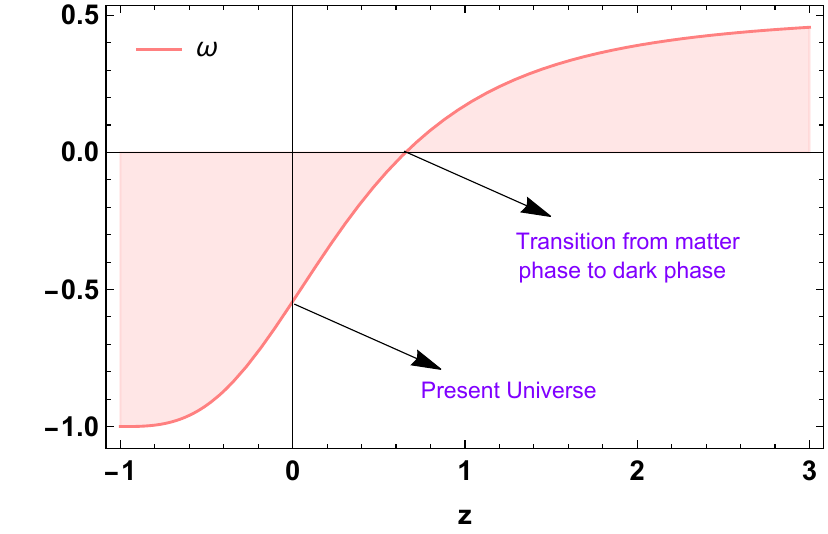}
	\caption{Above figure shows the behavior of $\omega$ versus $z$ with the  constraint values of the cosmological free parameters limited by $H(z)$ data set.}\label{M2EOS}
\end{figure}

The pressure evolution of our model's universe, as a function of redshift $z$, is illustrated in Fig. \ref{M2denp}. As $z$ approaches infinity ($p \rightarrow \infty$ as $z \rightarrow \infty$). Remarkably, the pressure becomes negative in the present era ($z=0$) and persists in the future ($z<0$), aligning with the anticipated behavior of dark energy. This result lends further credence to the efficacy of our $f(Q,T)$ gravity model in capturing the universe's accelerating expansion, providing a robust framework for understanding the cosmos.
\textit{\textbf{The equation of state parameter is}}

\begin{equation}
\omega =-1+ \frac{(\beta +2) (2 \beta +1) (n+1) (\dot{H}+2 \dot{H} n)}{\beta  (n+1) (\dot{H}+2 \dot{H} n)-3 (\beta +1) H^2 (2 n+1)}
\end{equation}
The redshift evolution of the equation of state (EoS) parameter $\omega$ is meticulously illustrated in Fig. \ref{M2EOS} , showcasing a fascinating trajectory that underscores the dynamic nature of cosmic expansion. Initially, the universe exhibits a stiff fluid phase, characterized by a positive EoS parameter ($\omega>0$), indicative of a decelerating expansion. As the universe evolves, the EoS parameter transitions to a quintessence phase ($\omega>-1$) at present, signaling a gradual shift towards accelerated expansion. Remarkably, the EoS parameter continues to evolve, asymptotically approaching the cosmological constant regime ($\omega=-1$) in the distant future, consistent with the predicted behavior of dark energy. The present-day value of the EoS parameter, $\omega_0 = -0.51318$, serves as a testament to the ongoing acceleration of the universe, reaffirming the validity of our $f(Q,T)$ gravity model in capturing the intricacies of cosmic evolution. This comprehensive evolutionary trajectory offers unparalleled insights into the dynamical interplay between matter, energy, and gravity, providing a refined understanding of the universe's expansion history.

\begin{figure}[H]
	\centering
	\includegraphics[scale=0.65]{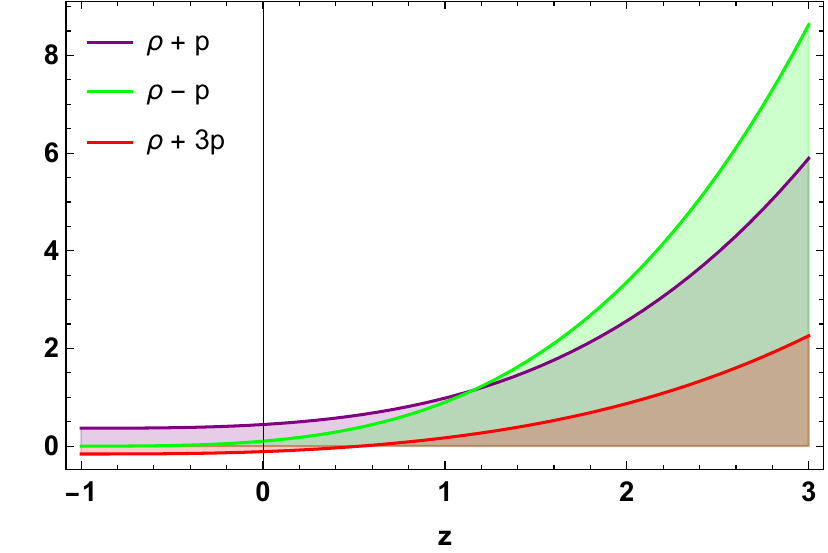}
	\caption{Above figure shows the behavior of $\rho + p, \rho - p$ and $\rho + 3p$ versus $z$ with the  constraint values of the cosmological free parameters limited by $H(z)$ data set.}\label{M2EC}
\end{figure}

\textit{\textbf{The energy conditions are}}\\
The explicit expressions for the Null Energy Condition (NEC), Dominant Energy Condition (DEC), and Strong Energy Condition (SEC) are derived as follows:

\begin{small}

\begin{equation}\label{eq:27} 
	\rho + p =  \frac{\alpha 2^{n+1} 3^n \left(2 n^2+3 n+1\right) H^{2n}\dot{H}}{\beta +1}       ,
\end{equation}

\begin{equation}\label{eq:27} 
		\rho - p =  -\frac{\alpha  2^{n+1} 3^n (2 n+1)  \left(\dot{H} n+\dot{H}+3 H^2\right)H^{2n}}{2 \beta +1}       ,
\end{equation}

\begin{equation}\label{eq:27} 
		\rho +3 p =  \frac{\alpha  2^{n+1} 3^n (2 n+1)  \left((5 \beta +3)  (n+1)\dot{H}+3 (\beta +1) H^2\right)H^{2n}}{(\beta +1) (2 \beta +1)}      .
\end{equation}


\end{small}
The dynamical evolution of the energy conditions is meticulously illustrated in Fig. \ref{M2EC}, leveraging the constrained values of the free parameters to elucidate the intricate behavior of these fundamental constraints. Our comprehensive analysis reveals that the Null Energy Condition (NEC), a crucial indicator of the universe's energetic landscape, exhibits a monotonic decreasing trend during the early universe, steadfastly remaining positive throughout this epoch but ultimately vanishing at late times, signaling a profound shift in the universe's energy dynamics. In stark contrast, the Dominant Energy Condition (DEC), a robust constraint on the universe's energy density, remains consistently positive throughout the entirety of cosmological evolution, without any violation, underscoring its enduring role in shaping the universe's large-scale structure. Meanwhile, the Strong Energy Condition (SEC), exhibits no violation during the early universe, but intriguingly violates at late times, hinting at a complex interplay between matter, energy, and gravity. These findings collectively underscore the dynamic and epoch-dependent nature of the energy conditions, highlighting their evolving role in sculpting the cosmological landscape and providing a nuanced understanding of the universe's energetic evolution.

\subsection{model-III}
\begin{figure}[H]
		\centering
		\includegraphics[scale=0.65]{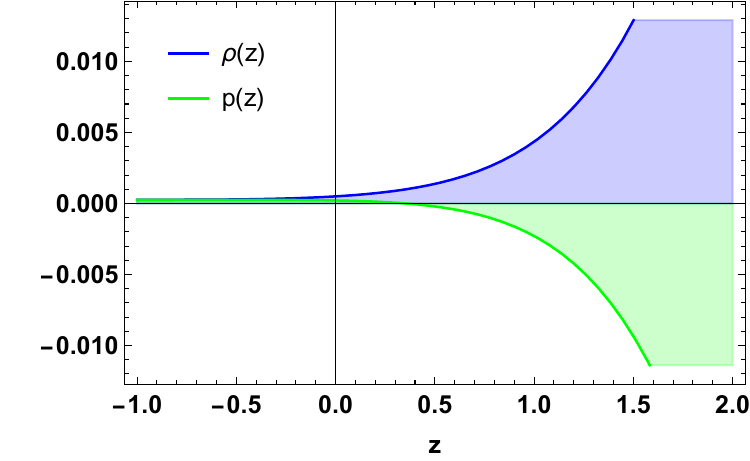}
		\caption{Above figure shows the behavior of $\rho$ and $p$ versus $z$ with the  constraint values of the cosmological free parameters limited by $H(z)$ data set.}\label{M3denp}
\end{figure}

Next, we explore a refined version of the $f(Q,T)$ gravity model, which assumes a quadratic functional form: $f(Q,T)=\alpha Q+ \beta Q^2+\gamma T$. This extended model incorporates a richer structure, enabling a more nuanced investigation of the non-metricity effects and their interplay with the matter-energy sector. The term $\alpha Q$ represents the linear contribution of non-metricity, while the term $\beta Q^2$ introduces a quadratic correction, allowing for a more subtle examination of the non-metricity's impact on cosmic evolution. Meanwhile, the term $\gamma T$ accounts for the influence of the matter-energy density, providing a comprehensive framework for understanding the intricate dynamics of the universe. By considering this refined model, we can probe the cosmological implications of non-metricity's quadratic corrections and their potential to address open questions in modern cosmology.\\
\textit{\textbf{The energy density  is}}

\begin{equation}
\rho =\frac{\alpha  \gamma  \dot{H}-3 H^2 (\alpha +\gamma  (\alpha -4 \beta  (\dot{H}+2 \dot{H})))-54 \beta  (\gamma +1) H^4}{(\gamma +1) (2 \gamma +1)}
\end{equation}
The , as described by the current $f(Q,T)$ gravity model, is mathematically represented by equation (36). The geometrical evolution of the energy density of the universe  with respect to redshift $z$ is vividly illustrated in Fig. \ref{M3denp}. Notably, as $z$ approaches infinity, the energy density diverges to infinity ($\rho \rightarrow \infty$ as $z \rightarrow \infty$), signifying that the universe was incredibly dense in its primordial stages. Conversely, as $z$ approaches -1, the energy density tends towards a small, yet positive value, indicating that the universe is presently undergoing expansion. This asymptotic behavior provides profound insights into the cosmological evolution of the universe, suggesting a smooth transition from an extremely dense, primordial state to a more diffuse, expansive phase. These findings offer valuable implications for our understanding of the universe's evolution within the framework of this $f(Q,T)$ gravity model.\\
\textit{\textbf{The isotropic pressure is}}
\begin{widetext}
\begin{equation}
p = \frac{(3 \gamma +2) \dot{H} \left(\alpha +12 \beta  H^2\right)+3 H^2 \left(\alpha  (\gamma +1)+18 \beta  (\gamma +1) H^2+8 \beta  (3 \gamma +2) \dot{H}\right)}{2 \gamma ^2+3 \gamma +1}
\end{equation}
\end{widetext}
The cosmological pressure evolution of our $f(Q,T)$ gravity model, as a function of redshift $z$, is depicted in Fig. \ref{M3denp}, showcasing a fascinating asymptotic behavior. As $z$ approaches infinity, the pressure diverges to infinity ($p \rightarrow \infty$ as $z \rightarrow \infty$), indicating a singular state in the distant past. Notably, the pressure undergoes a remarkable transition, becoming negative in the present era ($z=0$) and persisting in the future ($z<0$), consonant with the expected behavior of dark energy. This phenomenon is a hallmark of the accelerating expansion of the universe, which our model successfully captures. The emergence of negative pressure in the present and future epochs provides strong evidence for the viability of our $f(Q,T)$ gravity model, offering a robust theoretical framework for elucidating the cosmos's accelerating expansion and the underlying mechanisms driving this phenomenon.

\begin{figure}[H]
		\centering
		\includegraphics[scale=0.65]{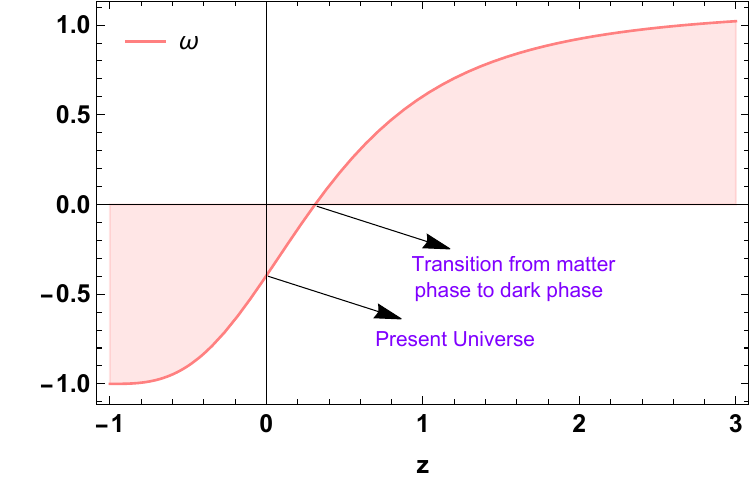}
		\caption{Above figure shows the behavior of $\rho$ and $p$ versus $z$ with the  constraint values of the cosmological free parameters limited by $H(z)$ data set.}\label{M3EOS}
\end{figure}

\begin{equation}
\omega = -1+\frac{(\gamma +2) (2 \gamma +1) \left(\alpha  \dot{H}+12 \beta  H^2 (\dot{H}+2 \dot{H})\right)}{\alpha  \gamma  \dot{H}-3 H^2 (\alpha +\gamma  (\alpha -4 \beta  (\dot{H}+2 \dot{H})))-54 \beta  (\gamma +1) H^4}
\end{equation}
The redshift evolution of the equation of state (EoS) parameter $\omega$ is meticulously depicted in Fig. \ref{M3EOS}, revealing a captivating trajectory that underscores the dynamic and evolving nature of cosmic expansion. Initially, the universe exhibits a stiff fluid phase, characterized by a positive EoS parameter $(\omega > 0)$, indicative of a decelerating expansion regime. As the universe undergoes evolution, the EoS parameter undergoes a transition to a quintessence phase $(\omega > -1)$ at present, signaling a gradual shift towards an accelerated expansion paradigm. Remarkably, the EoS parameter continues to evolve, asymptotically approaching the cosmological constant regime $(\omega = -1)$ in the distant future, consistent with the predicted behavior of dark energy. The present-day value of the EoS parameter, $\omega_0 = -0.48118$, serves as a testament to the ongoing acceleration of the universe, reaffirming the validity and efficacy of our $f(Q,T)$ gravity model in capturing the intricacies and nuances of cosmic evolution. This comprehensive evolutionary trajectory offers unparalleled insights into the dynamical interplay between matter, energy, and gravity, providing a refined and nuanced understanding of the universe's expansion history, and underscoring the complex and evolving nature of the cosmos.
\textit{\textbf{The energy conditions are}}\\
The explicit expressions for the Null Energy Condition (NEC), Dominant Energy Condition (DEC), and Strong Energy Condition (SEC) are derived as follows:
\begin{small}
\begin{equation}\label{eq:27}
\rho + p=  \frac{2  \left(\alpha +36 \beta  H^2\right)\dot{H}}{\gamma +1}       ,
\end{equation}
\begin{equation}\label{eq:27}
		\rho - p =  -\frac{2 \left(\alpha  \dot{H}+36 \beta  \dot{H} H^2+54 \beta  H^4+3 \alpha  H^2\right)}{2 \gamma +1}       ,
\end{equation}
\begin{equation}\label{eq:27}
		\rho +3 p = \frac{2 (5 \gamma +3)  \left(\alpha +36 \beta  H^2\right)\dot{H}+6 (\gamma +1)  \left(\alpha +18 \beta  H^2\right)H^2}{2 \gamma ^2+3 \gamma +1}       .
\end{equation}
\end{small}

\begin{figure}[H]
	\centering
	\includegraphics[scale=0.6]{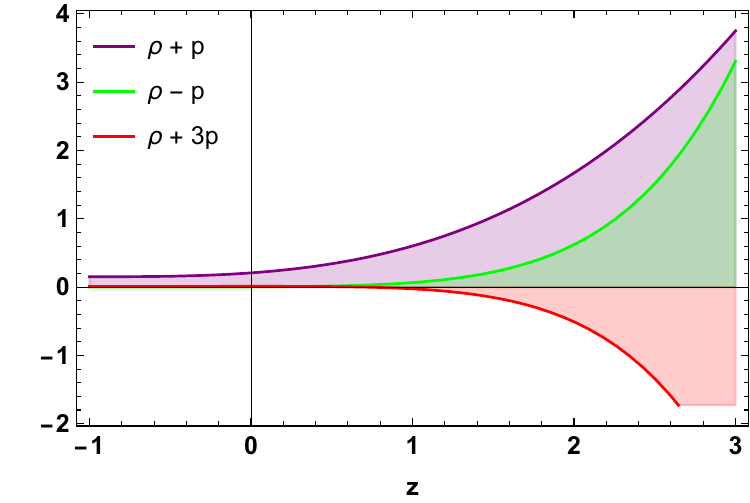}
	\caption{Above figure shows the behavior of $\rho + p, \rho - p$ and $\rho + 3p$ versus $z$ with the  constraint values of the cosmological free parameters limited by $H(z)$ data set.}\label{M3EC}
\end{figure}

The dynamical evolution of the energy conditions is meticulously illustrated in Fig. \ref{M3EC}, leveraging the constrained values of the free parameters to elucidate the intricate behavior of these fundamental constraints. Our comprehensive analysis reveals that the Null Energy Condition (NEC) exhibits a monotonic decreasing during the early universe, steadfastly remaining positive throughout this epoch but ultimately vanishing at late times. In contrast, the Dominant Energy Condition (DEC) remains consistently positive, without any violation, throughout the entireness of cosmological evolution. Meanwhile, the Strong Energy Condition (SEC) exhibits no violation during the early universe, but intriguingly violates at late times. These findings collectively underscore the dynamic and epoch-dependent nature of the energy conditions for the understanding of the universe's energetic evolution.

\section{Conclusion}
In this manuscript, we investigate the dynamical behavior of perfect fluid sources within the theoretical framework of modified gravity, specifically $f(Q,T)$ gravity, situated within the Friedmann-Lemaître-Robertson-Walker (FLRW) cosmological paradigm. To derive exact solutions to the field equations, we adopt a parameterized form of the Hubble parameter, $H(z)$. Our analysis reveals intriguing insights into the physical and geometrical properties of the model, yielding the following key results:\\
\textbf{\textit{Cosmic Kinematics:} }\\
The transition from a decelerating to an accelerating expansion phase is a crucial aspect of understanding the universe's dynamical evolution. Initially, the universe's expansion was decelerating due to the strong gravitational attraction between matter and radiation, characterized by a positive deceleration parameter (q). However, as the universe expanded and matter became increasingly dispersed, the gravitational force weakened, leading to a shift in cosmic dynamics. The universe has now entered a phase of accelerated expansion, marked by a negative deceleration parameter (q), which provides a quantitative measure of this acceleration. The deceleration parameter (q) serves as a diagnostic tool, distinguishing between two phases of cosmic evolution: a positive value indicates deceleration, while a negative value signifies acceleration. Our analysis yields a present-day deceleration parameter value of q0, which is consistent with recent observational constraints, q0 = -0.528+0.092 -0.088, thereby providing insights into the current state of cosmic expansion and the universe's evolutionary trajectory.\\
Next we discussed the State-finder parameters. As, the significance of dark energy in driving the accelerating cosmic expansion is well-established. In recent decades, there has been a surge of interest in elucidating the origin and fundamental properties of dark energy, leading to the development of numerous dark energy models These parameters provide a robust means of distinguishing between various dark energy models, enabling a deeper understanding of the underlying mechanisms driving the cosmic acceleration.\\
Furthermore, we explored the $O_m(z)$ diagnostic, a versatile and powerful tool for investigating the accelerated expansion of the Universe, providing a complementary approach to conventional methods. This diagnostic boasts the capability to discern a wide array of dark energy models, encompassing quintessence and phantom scenarios. Notably, the slope of the $O_m(z)$ function serves as a discriminatory metric, distinguishing between disparate dark energy models: a positive slope is indicative of a phantom regime, characterized by an equation of state parameter $\omega < -1$, whereas a negative slope typifies a quintessence domain, marked by $\omega > -1$. This diagnostic tool offers a valuable means of constraining dark energy models, enabling researchers to elucidate the underlying mechanisms driving the observed cosmic acceleration.\\
\textbf{\textit{Cosmological Implications of $f(Q,T)$ Gravity Models:}} \\
Through a comprehensive examination of diverse functional forms of $f(Q,T)$ gravity models, specifically three distinct formulations, we investigate the physical parameters that underlie the model's framework. By scrutinizing these parameters, we can decipher their role in shaping the cosmological evolution and uncover the intrinsic characteristics of the model. This analysis enables us to elucidate the physical implications of the model, providing valuable insights into its behavior and consequences for our understanding of the universe. By exploring the physical parameters that govern the model, we can gain a deeper understanding of the underlying mechanisms driving the observed cosmological phenomena.\\
In all three $f(Q,T)$ gravity models, the energy density ($\rho$) exhibits a monotonic increase with redshift, asymptotically diverging to infinity as $z$ approaches infinity ($\rho \rightarrow \infty$ as $z \rightarrow \infty$). This behavior indicates that the universe was characterized by an extremely high density in its primordial stages. Moreover, as $z$ approaches -1, the energy density converges towards a small positive value, suggesting an expanding universe. These findings offer valuable insights into the cosmological evolution of the universe within the framework of this $f(Q,T)$ gravity model. Notably, the pressure becomes negative in the present era ($z=0$) and remains so in the future ($z<0$), consistent with the expected behavior of dark energy as predicted by astronomical observations. The negative pressure is a key indicator of dark energy's presence, which is thought to be responsible for the accelerating expansion of the universe.\\
The ultimate nature of dark energy (DE) is often characterized by the Equation of State (EoS) parameter, which quantifies the relationship between spatially homogeneous pressure and energy density. Recent cosmological investigations have highlighted the significance of the EoS parameter, $\omega < -1/3$, as a necessary condition for rapid cosmic acceleration. In all three models, the evolutionary trajectory of the EoS parameter ($\omega$) exhibits a transition from a positive value to a quintessence phase, where $\omega > -1$ at present, and asymptotically approaches $\omega = -1$ in the future. This evolution implies a transition from a decelerating to an accelerating phase of cosmic expansion. Notably, the present-day values of the EoS parameter, $\omega_0 = -0.58318, -0.51318, -0.48118$, confirm that the universe is currently undergoing acceleration. These findings provide valuable insights into the dynamical evolution of the universe within the framework of our $f(Q,T)$ gravity model, offering a deeper understanding of the underlying mechanisms driving the observed cosmic acceleration.\\
The energy conditions, a fundamental concept in modern cosmology, play a pivotal role in predicting cosmic acceleration. Extensive research has been dedicated to exploring various forms of energy conditions, including the null energy condition (NEC), dominant energy condition (DEC), and strong energy condition (SEC). In this study, we investigate these well-established energy conditions to assess the viability of our model in explaining cosmic acceleration. Our analysis reveals that, in all three models, the NEC ($\rho + p > 0$) exhibits a monotonic decrease during the early universe, remaining steadfastly positive throughout this epoch but ultimately vanishing at late times. In contrast, the DEC ($\rho - p > 0$) remains consistently positive throughout the entire cosmological evolution, without any violation. Meanwhile, the SEC ($\rho + 3p > 0$) exhibits no violation during the early universe but intriguingly violates at late times. These findings collectively underscore the dynamic and epoch-dependent nature of the energy conditions, highlighting their significance in understanding the universe's energetic evolution and the mechanisms driving cosmic acceleration.\\
 Cosmography, in the context of modified gravity, particularly f(Q,T) gravity, plays a crucial role in understanding the universe's expansion and the behavior of cosmological parameters without assuming a specific model of gravity.
The Hubble parameter $H(z)$ is crucial in cosmography as it describes the expansion rate of the universe at different redshifts $z$. Specific parameterizations of $H(z)$ allow cosmologists to test various models and constrain cosmological parameters. In $f(Q,T)$ gravity, the choice of $H(z)$ can help in understanding deviations from standard cosmology and provide insights into the effects of non-metricity and the energy-momentum trace on cosmic expansion. In summary, cosmography in $f(Q,T)$ modified gravity is essential for providing a model-independent framework to study the universe's expansion, constraining modified gravity theories, and offering insights into the nature of cosmic acceleration and the fundamental properties of the universe.

\section*{Declaration of competing interest}
The authors declare that they have no known competing financial interests or personal relationships that could have appeared to influence the work reported in this paper.

\section*{Data availability}
No data was used for the research described in the article.

\section*{Acknowledgments}
The IUCAA, Pune, India, is acknowledged by the authors (S. H. Shekhi, A. Pradhan) for giving the facility through the Visiting Associateship programmes.

\end{document}